\documentclass[twocolumn,showpacs,preprintnumbers,amsmath,amssymb]{revtex4}
\usepackage{graphicx}
\usepackage{amsmath}
\usepackage{amsfonts}
\newcommand{\be}{\begin{equation}}
\newcommand{\en}{\end{equation}}
\newcommand{\no}{\nonumber}
\newcommand{\fr}[2]{\frac{#1}{#2}}
\newcommand{\q}{\quad}
\begin{document}
\title{Extremal Channels for a genaral Quantum system}
\author{Kuldeep Dixit}
\email{kuldeep@physics.utexas.edu}
\author{E. C. G. Sudarshan}
\affiliation{Department of Physics,
        University of Texas,
        Austin, TX 78712.}

\begin{abstract}

Quantum channels can be mathematically represented as completely
positive   trace-preserving maps that act on a density matrix. A
general quantum channel can be written as a convex sum of
`extremal' channels. We show that for an $N$-level system, the
extremal channel can be characterized in terms of $N^2$-$N$ real
parameters coupled with rotations. We give a representation for
$N$= 2, 3, 4.
\end{abstract}
\maketitle

\section{Introduction}

A completely positive map acting on an N-level open quantum system
can be characterized in terms of $C$ matrices as \cite{1,2,3,4}:

\begin{equation}
\rho'= B(\rho)=\sum_i C_i \: \rho \: C_i^\dag.
\end{equation}

The above form automatically satisfies the Hermiticity and
positive semi-definiteness conditions required for the resulting
density matrix. Since the trace of $\rho$ must also be preserved,
C matrices satisfy the additional condition given by: \be \sum_i
C_i^ \dag C_i = I. \label{Trace} \en

In terms of open quantum system dynamics, Completely positive maps
arise from the unitary evolution of a larger system that comprises
the system and the environment starting out in a simply separable
state, i. e.

\begin{equation}
\rho'=Tr_{env}[U \: (\rho_{sys}\otimes \rho_{env}) \: U^\dag].
\label{env}
\end{equation}

The most general map acting on $\rho$ can be expressed in terms of
$N^2$ such $C$ matrices. However, a general map can be written as
a convex sum of `extramal' maps \cite{5,6}. For example, Unitary
evolution of a density matrix is equivalent to having only one C
matrix. Though trivial, it turns out that such maps play an
important part in center preserving (Unital) maps because in two
and four dimensional quantum systems a general Unital map can be
written as a convex sum of unitary maps \cite{7,8}.

Physically, extremal maps arise if the environment starts out in a
pure state in eq. (\ref{env}). In addition to satisfying eq.
(\ref{Trace}), the C matrices for the extremal map can be written
in a manifestly trace orthonormal form, i. e., for $i \ne j$

\be
 Tr [C_i^\dag C_j]=0.
\label{tr orth}
 \en
Based on this condition it can be easily seen that a general
extremal map can only have a maximum of N such C matrices. Ruskai
et. al. \cite{7} have explored extremal maps for a qubit case in
detail and have found that apart from rotations, the extremal map
can be expressed in terms of two independent parameters. We
generalize this result and give a representation of the map for
quantum systems with up to four levels.

\section{Theorem 1} Apart from rotations, a general extremal map acting on an N level system can be
expressed in terms of $N^2-N$ real parameters.

\textbf{Proof:} Let's start with N C matrices and write the
singular value decomposition of each one as \be C_i=U_iD_iV_i. \en
Here U and V are unitary matrices and D is a real diagonal matrix
with positive entries. First we show that the V's are just
permutation matrices and can be dropped from parameter counting.
We provide a proof by induction. Let's start with a two level
system. Equation (\ref{Trace}) imposes the condition:

\be V_1^\dag D_1^2 V_1 + V_2^\dag D_2^2 V_2=I. \label{qubit1} \en
Multiplying the whole equation on the left with $V_1$ and on the
right with $V_1^\dag$ gives us

\be D_1^2 + {V'}_2^\dag D_2^2 V'_2=I \en

or \be  {V'_2}^\dag D_2^2 V'_2=I-D_1^2 . \en

Since the right hand side is diagonal and the
operation on the left hand side does not change the eigenvalues of
$D_2^2$, the only effect that $V'_2$ has is to permute the
eigenvalues of $D_2^2$. In other words we can absorb the $V's$ in D itself. For a three level system we have:

 \be V_1^\dag D_1^2 V_1 + V_2^\dag D_2^2
V_2 + V_3^\dag D_3^2 V_3=I. \en
Once again, this results into

\be {V'_1}^\dag D_1^2 V'_1 + {V'_2}^\dag D_2^2 V'_2=I-D_3^2. \en
Since $I-D_3^2$ is a positive matrix, we can define the square
root of its multiplicative inverse. Let's call it $M$.
Multiplication of $M$ on both sides gives us

\be M {V'_1}^\dag D_1^2 V'_1 M + M {V'_2}^\dag D_2^2 V'_2M =I. \en

Since $M {V'_1}^\dag D_1^2 V'_1 M$ is both positive and Hermitian,
we can write it as ${V''_1}^\dag {D'_1}^2 V''_1$. Using this the
above equation is reduced to our already worked case of equation
(\ref{qubit1}). Thus we have shown that qutrit (and thus by induction all
higher level systems) can be reduced to the qubit case.

We have shown that V's appear trivially in C. As the next step let's examine U's.
The trace orthogonality equation (\ref{tr orth}) results into:

\be
 Tr [D_j D_i U_i^\dag U_j]=0.
\en
Since $D_j D_i$ is a diagonal matrix, the above equation is
satisfied if we let $U \equiv U_i^\dag U_j$ have only non-diagonal
entries. Such a U can be trivially constructed and we give
examples for up to four dimensional quantum systems.

As seen from above each of the C matrices can be characterized in
terms of just D matrices. Since for the most general case there are N C matrices and thus N of these diagonal matrices (requiring $N^2$ real parameters) and there is only one
constraint on them, namely the completeness condition (which
imposes N constraints), the extremal map will have only $N^2-N$
independent parameters. In the following section we give some
simple examples.

\section{Examples of extremal maps}

For the N=2 case the simple choice of matrices is given by:

\be D_1=diag(a,b), \q D_2=\sqrt{I-D_1^2}, \q U_1=I, \q
U_2=\sigma_1. \en

The geometry of the map can be made more transparent by putting

\begin{eqnarray}
a=\mu_0 + \mu_3 \no \\
b=\mu_0 - \mu_3 \no \\
\mu_0=\fr{1}{4} \sqrt{1+\nu_1+\nu_2+\nu_3} \no \\
\mu_3=\fr{1}{4} \sqrt{1+\nu_3-\nu_1-\nu_2}.
\end{eqnarray}

Geometrically the map is a depolorizing map (see \cite{8}) coupled
with translation, i. e. the map shrinks $\sigma_i$ polarization by
a factor of $\nu_i$ and translates the resulting Bloch ellipsoid
along $\sigma_3$ by an amount

\be t_3= \sqrt{(1-\nu_3)^2-(\nu_1-\nu_2)^2}. \en

The completeness condition gives the relation

\be \nu_3=\nu_1 \nu_2. \en

Thus the shrinking along $\sigma_3$ is a product of the shrinkages
along the other two directions. The independent parameters of the
map are $\nu_1$ and $\nu_2$.

\begin{figure}
  \includegraphics[width=6 cm]{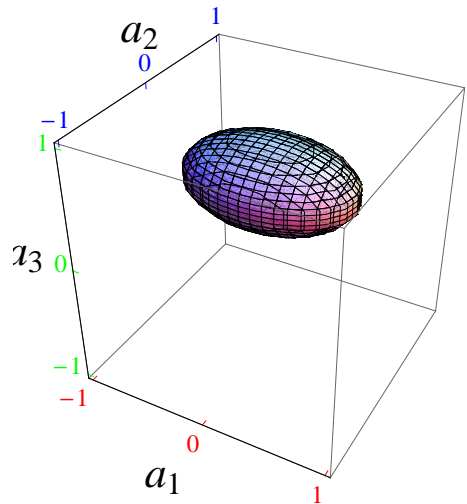}\\
  \caption{"(color online)" Extremal map acting on a qubit compresses the Bloch sphere along the three polarizations by {$\nu_1$,  $\nu_2$, $\nu_1 \nu_2$} and translates the resulting ellipsoid by $\sqrt{(1-\nu_1 \nu_2)^2-(\nu_1-\nu_2)^2}$ along the $\sigma_3$ direction. }\label{qubit}
\end{figure}

For $N=3$ case the geometry of the map is much more complicated.
However the C matrices can be easily constructed as

\begin{eqnarray}
D_1=diag(a,b,c), \q D_2=diag(d,e,f), \q D_3=\sqrt{I-D_1^2-D_2^2}
\no
\\
U_1=\left(%
\begin{array}{ccc}
  1 & 0 & 0 \\
  0 & 0 & 1 \\
  0 & 1 & 0 \\
\end{array}%
\right), \q U_2=\left(%
\begin{array}{ccc}
  0 & 0 & 1 \\
  0 & 1 & 0 \\
  1 & 0 & 0 \\
\end{array}%
\right), \q U_3=\left(%
\begin{array}{ccc}
  0 & 1 & 0 \\
  1 & 0 & 0 \\
  0 & 0 & 1 \\
\end{array}%
\right).
\end{eqnarray}

Finally, for $N=4$ the U matrices are given by:

\be U_1=I, \q U_2=1\otimes\sigma_1, \q U_3= \sigma_1 \otimes 1, \q
U_4=\sigma_1 \otimes \sigma_1. \en

\section{Conclusion}
We have shown how to represent an extremal map for dimensions N = 2,
3, 4. N=4 level case is particularly important since it allows for
entanglement between two qubit subsystems. The generalization to higher
level systems is easily accomplished. In each case the map can be
written in terms of $N^2-N$ real parameters. For N=2 case the
parameters present a very intuitive geometrical picture, namely
that the two parameters are the compression coefficients along two
axes of the block sphere. The compression along the third
direction is just the product of these two compressions. For
higher level systems the geometry becomes more difficult, however
it would be very insightful to see if similar relations exist
between the compression coefficients.


\begin{thebibliography}{1}
\bibitem{1} E. C. G. Sudarshan, P. M. Mathews, and J. Rau, Phys. Rev. 121,
920 (1961).

\bibitem{2} T. F. Jordan and E. C. G. Sudarshan, J. Math. Phys. \textbf{2}, 772
(1961).

\bibitem{3} E. St{\o}rmer, Acta Math. \textbf{110}, 233 (1963).

\bibitem{4} M. D. Choi, Can. J. Math. \textbf{24}, 520 (1972).

\bibitem{5} E. C. G. Sudarshan,  V. Gorini, Comm.
Math. Phys. 46, 43 (1976).

\bibitem{6} E. C. G. Sudarshan, From
SU(3) to Gravity (Festschrift in honor of Yuval Ne'eman), E.
Gotsman and G. Tauber (eds.), Cambridge University Press (1986),
p. 433.

\bibitem{7}Mary Beth Ruskai, Stanislaw Szarek, and Elisabeth Werner, Linear
Algebr. Appl. 347, 159 (2002).

\bibitem{8} K. Dixit and E. C. G. Sudarshan, Phys. Rev. A 78, 032308
(2008).

\end{thebibliography}
\end{document}